\documentclass[a4paper,twoside]{article}
\newcommand{\affil}[1]{$^{\rm #1}$}
\usepackage[authoryear]{natbib}
\bibpunct{(}{)}{;}{a}{}{,}
\usepackage{graphicx}
\date{} 
\title{\large\bf\flushleft Probing the Origins of Voids in the Distribution of Galaxies}
\author{\parbox{\textwidth}{\flushleft
\vspace{-0.5cm}
{\it Louise M. Ord \affil{A,B,E}, Martin Kunz \affil{B,C}, Hugues Mathis \affil{D}, and Joseph Silk \affil{D}}\\
\vspace{0.4cm}
{\small \affil{A}\,Department of Astrophysics \& Optics, School of Physics, University of New South Wales, Sydney, NSW 2052, Australia}\\
{\small \affil{B}\,Astronomy Centre, University of Sussex, Brighton BN1 9QJ, UK}\\
{\small \affil{C}\,Theoretical Physics, University of Geneva, 1211 Geneva 4, Switzerland}\\
{\small \affil{D}\,Astrophysics, University of Oxford, Denys Wilkinson Building, Keble Road, Oxford OX1 3RH, UK}\\
{\small \affil{E}\,Email: louise@phys.unsw.edu.au}}}
\begin{document}
\twocolumn[
\begin{changemargin}{.8cm}{.5cm}
\begin{minipage}{.9\textwidth}
\vspace{-1cm}
\maketitle
%
%
\small{\bf Abstract:} If the voids that we see today in the
distribution of galaxies existed at recombination, they will leave an
imprint on the cosmic microwave background (CMB).  On the other hand,
if these voids formed much later, their effect on the CMB will be
negligible and will not be observed with the current generation of
experiments. In this paper presented at the 2004 Annual Scientific
Meeting of the Astronomical Society of Australia, we discuss our
ongoing investigations into voids of primordial origin. We show that
if voids in the cold dark matter distribution existed at the epoch of
decoupling, they could contribute significantly to the apparent rise
in CMB power on small scales detected by the Cosmic Background Imager
(CBI) Deep Field. Here we present our improved method for predicting
the effects of primordial voids on the CMB in which we treat a void as
an external source in the cold dark matter (CDM) distribution
employing a Boltzmann solver. Our improved predictions include the
effects of a cosmological constant ($\Lambda$) and acoustic
oscillations generated by voids at early times.  We find that models
with relatively large voids on the last scattering surface predict too
much CMB power in an Einstein--de Sitter background cosmology but
could be consistent with the current CMB observations in a
$\Lambda$CDM universe.

\medskip
\medskip
\end{minipage}
\end{changemargin}
]
\small

\section{Introduction}
Analyses of surveys such as the 2-degree Field Galaxy Redshift Survey
and the Sloan Digital Sky Survey indicate large volumes of relatively
empty space, or voids, in the distribution of galaxies
\citep[e.g.][]{hoy02,hoy04}. In the standard hierarchical model of
structure formation, gravitational clustering is responsible for
emptying these voids of mass and galaxies \citep{pee89}. However,
standard cold dark matter (CDM) simulations predict significant clumps
of substructure within voids that that should be capable of developing
into observable bound objects, such as dwarf galaxies, that are not
apparent in the surveys \citep{dek86,hof92,sha04}.  \citet{pee01}
gives an in-depth discussion of the contradictions of this prediction
with observation.  He believes that the inability of the CDM models to
produce the observed voids constitutes a true crisis for these models.

It has been argued that the apparent suppression of void object
formation may be addressed when more sophisticated models for the
complex effects of baryons on galaxy formation are developed
\citep[e.g.][]{van00}. Others believe that alternatives to CDM, such
as warm dark matter (WDM) \citep{hog00,som01} or self-interacting dark
matter \citep{spe00}, may provide a solution. The higher temperature
of WDM particles causes them to move faster than particles of CDM.
This motion might enable WDM particles to resist congregating to form
observable galaxies within voids. However, \citet{bar01} note that the
material's resistance to clumping might delay the epoch of quasar
formation and the model is ruled out if reionisation occurred before
$z\sim10$ as indicated by the recent WMAP cosmic microwave background
(CMB) data \citep{spe03}.

Self-interacting dark matter may explain the relative dearth of dwarf
galaxies. If there were interparticle collisions, the halo of dark
matter surrounding a large galaxy would interact with the halos of
nearby dwarf galaxies, stripping the dwarves of their gas and stars
more rapidly than in the standard CDM theory. However, Chandra X-ray
Observatory data indicates that the density of the dark matter is
greater the closer it is to the center of the cluster~\citep{ara02},
in contradiction with the simplest self-interacting dark matter models
which predict particle collisions to occur most frequently in more
populated regions and act to spread out galactic cores and reduce
their density.

The standard CDM model of structure formation remains extremely
successful at matching the observations on large scales such as the
properties of galaxy clusters, the statistics of the Lyman-$\alpha$
forest and the temperature anisotropy of the CMB.  However, the lack
of observable dwarf galaxies within voids as predicted by the
simulations can not be overlooked. Additionally, recent deep field
observations from the Cosmic Background Imager (CBI) \citet{mas03}
show excess power in the CMB over the standard model on small angular
scales, $\ell > 2000$. 

It may be possible to explain both these observations by postulating
the presence of a network of voids originating from primordial bubbles
of true vacuum that nucleated during inflation \citep{la91,lid91}. In
this scenario, the first bubbles to nucleate are stretched by the
remaining inflation to cosmological scales. The largest voids may have
had insufficient time to thermalise before decoupling and may persist
to the present day.  Such primordial voids are predicted to produce a
measurable contribution to the CMB on small-scales (see \citet{gri03}
and references therein). If voids in the matter distribution were
present prior to the onset of gravitational collapse, the formation of
bound objects within voids would be suppressed.

In this paper, we discuss the results of \citet{gri03} in which we
develop a general method to approximate the void contribution to the
CMB that allows the creation of maps and enables us to consider an
arbitrary distribution of void sizes. We show that if the voids that
we see in galaxy surveys today existed at the epoch of decoupling,
they could contribute significant additional power to the CMB angular
power spectrum between $2000 < \ell < 3000$. Further to \citet{gri03},
we show how we improve our predictions. We include the effects of the
cosmological constant as well as oscillations in the matter-radiation
fluid that may be generated by primordial voids on scales up to the
sound horizon \citep{bac00} and present our method for constraining
the parameters of this inflationary model of void production.

\section{First Approximation}
\subsection{Void parameters}
In \citet{gri03} we model the voids seen today as spherical
underdensities of $\delta \rho/\rho = -1$.  As a primordial void
bubble nucleates and is stretched by inflation it will sweep up the
surrounding energy density to form a thin boundary wall.  At the end
of the inflationary epoch, any voids that persist will be fully
compensated. This means that the overall cosmology is that of the
background universe, since a compensated void does not distort
space-time outside of itself.  This is an extremely important
property, as it allows us to place many voids into a universe without
the worry that they might influence each other.  As a first
approximation we take the background universe to be an Einstein--de
Sitter (EdS) cosmology. \citet{mae83} and \citet{ber85} use
conservation of momentum and energy respectively to show that
compensated voids in an EdS background cosmology will increase in
radius $r_v$ between the onset of the gravitational collapse of matter
at equality and the present day such that
\begin{equation}
\label{e:vexpand}
r_v(\eta) \propto \eta^{\beta} \,,
\end{equation}
with $\beta \approx 0.39$ and where $\eta$ is conformal time.

Motivated by the extended inflationary scenario \citep[][see Kolb 1991
for a review]{las89}, we assume a power-law distribution of bubble
sizes greater than a given radius $r$ of the form,
\begin{equation}
N_B(>r) \propto r^{-\alpha} . \label{eqvoidsize}
\end{equation}
Typically, extended inflation is implemented within the framework of a
Jordan-Brans-Dicke theory \citep{bra61}. In this case, the exponent
$\alpha$ is directly related to the gravitational coupling $\omega$ of
the scalar field that drives inflation,
\begin{equation}
\alpha = 3 + \frac{4}{\omega + 1/2} .
\end{equation}
Values of $\omega>3500$ are required by solar system experiments (Will
2001). We take the limit of large $\omega$, leading to a spectrum of
void sizes with $\alpha = 3$.

\citet{lid91} show that a distribution of void sizes that predicts the
existence of arbitrarily large inflationary voids will cause
significant effects on the CMB that contradict observations.  We
assume that the mechanism creating the voids imposes an upper cut-off
on the size distribution. A possible mechanism for this cut-off could
be that the tunneling probability of inflationary bubbles is modulated
through the coupling to another field. 

As well as avoiding the well known problems associated with
arbitrarily large voids existing at the epoch of decoupling, this
assumption allows us to match the observed upper limit on void sizes
from the galaxy redshift surveys. Since the results of void finder
algorithms in the literature include those obtained from incomplete
galaxy surveys, we choose to take the average maximum void size that
is found, $r_{\rm max} = 25$ Mpc$/h$ \citep[e.g.][]{hoy02,hoy04}. An
analysis is also in progress in which this parameter is kept free so
that it may be constrained by the observations. The minimal present
void size is also chosen to agree with redshift surveys, $r_{\rm
min}=10$ Mpc$/h$.

We normalise the distribution by requiring that the total number of
voids satisfies the observed fraction of the universe filled with
voids today, $F_v$.  Redshift surveys indicate that approximately 40\%
of the fractional volume of the universe is in the form of voids of
underdensity $\delta \rho/\rho < -0.9$ \citep[e.g.][]{hoy02,hoy04},
ie. $F_v \approx 0.4$. The positions of the voids are then assigned
randomly, making sure that they do not overlap. In order to speed up
this process, we consider only a $10^\circ$ cone. This limits our
analysis to $\ell > 100$, which is satisfactory for our purpose since
the main contribution from voids is on much smaller scales.

\subsection{Stepping through the void network}\label{sec:vstep}
Refer to \citet{gri03} for a more detailed description of our
methodology.  We ray trace photon paths from us to the last scattering
surface (LSS) for the 10$^\circ$ cone in steps of 1'.  Each void in
the present day distribution that is intersected by the photon path is
evolved back in time according to equation~(\ref{e:vexpand}) to
determine whether the photon encounters the void.

If a photon intersects a void between us and the LSS, we compute the
Rees--Sciama (RS) (1968) effect due to the deviation in the redshift
of the photon as it passes through the expanding void and the lensing
effect due to the deviation in its path. If a photon intersects a void
on the LSS, we calculate the Sachs--Wolfe (SW) (1967) effect due to
the photon originating from within the underdensity. We take into
account the finite thickness of the LSS, which suppresses the SW
effect for small voids, by averaging the contribution from a number of
photons originating from a LSS of mean redshift 1100 and standard
deviation in redshift 80.  We also calculate the partial RS effect
that arises due to the expansion of the void on the LSS as the photon
leaves it.

\begin{figure}[t]
\begin{center}
\includegraphics[scale=0.4, angle=0]{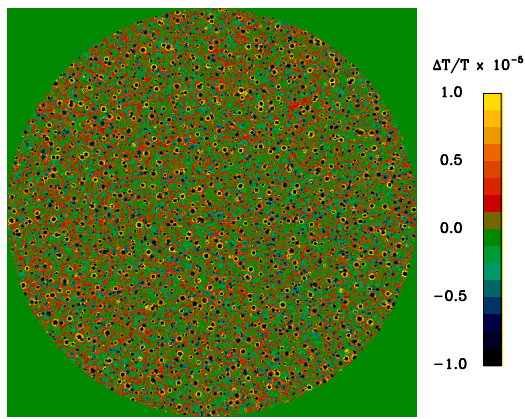}
\caption{The map of the temperature fluctuations on the surface of
last scattering from the fiducial EdS void model considered. The
parameters of this model are given by $\alpha = 3$, $r_{\rm max}=25$
Mpc$/h$ and $F_v=0.4$.}\label{figmap}
\end{center}
\end{figure}
\begin{figure}[t]
\begin{center}
\includegraphics[scale=0.4, angle=0]{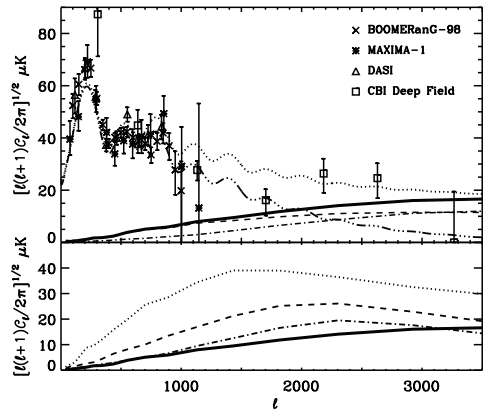}
\caption{Top: The CMB anisotropies produced by the fiducial EdS void
model (solid line) compared to the primary $\Lambda$CDM CMB
anisotropies (dashed-triple-dotted). Also plotted are the sum of
primary and void contributions (dotted) as well as the fluctuations
induced purely by voids on the last scattering surface (dashed) and by
those between last scattering and today (dashed-dotted).  We show the
``standard'' cosmological concordance model: of course a combined
analysis of primary and void--induced fluctuations would select a
different cosmology for the primary contribution.  Bottom: Example
models depicting a range of void contributions to the CMB
fluctuations. The models plotted are $\alpha = 3$, $r_{\rm max}=25$
Mpc$/h$ and $F_v=0.4$ (solid line), $\alpha = 3$, $r_{\rm max}=40$
Mpc$/h$ and $F_v=0.4$ (dotted), $\alpha = 3$, $r_{\rm max}=40$ Mpc$/h$
and $F_v=0.2$ (dashed) and $\alpha = 6$, $r_{\rm max}=40$ Mpc$/h$ and
$F_v=0.4$ (dashed-dotted).}\label{figcl}
\end{center}
\end{figure}

Once the photon has reached the last scattering surface, we know the
variation of its temperature as well as its position on the LSS and
can create a temperature map (figure \ref{figmap}).  We then use a
flat sky approximation \citep{whi99,das02} to obtain the $C_\ell$
spectrum of the anisotropies (see figure \ref{figcl}). We point out
that primordial void parameters are still poorly constrained by both
observation and theory.  The bottom panel of figure~\ref{figcl} shows
a few further example models.

For a power-law size distribution (as motivated by the inflationary
scenario), large voids become rarer as $\alpha$ is increased.
Therefore, since void analyses of redshift surveys only sample a
fraction of the volume of the universe, there may exist voids of
larger $r_{\rm max}$ than currently observed.  Models with high
$r_{\rm max}$ voids in an EdS background cosmology tend to predict too
much power on scales $\ell \approx 1000$. However, if we take
inflationary models with $\alpha > 6$, as motivated by
eg. \citet{occ94}, then the peak moves to larger $\ell$ and the total
power drops. The filling fraction mainly adjusts the overall power.


\section{Improved Prediction}
\subsection{Void Evolution in a $\Lambda$CDM background cosmology}
\begin{figure}[t]
\begin{center}
\includegraphics[scale=0.3, angle=0]{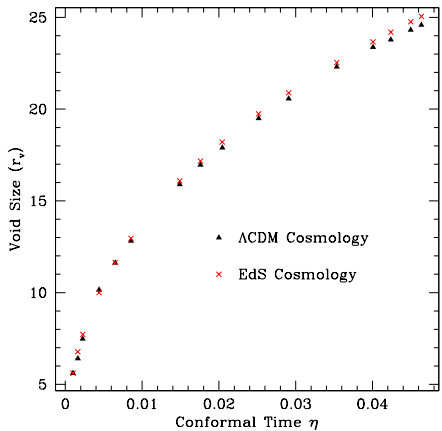}
\caption{The simulated size evolution for a 25 Mpc$/h$ void in a
$\Lambda$CDM background cosmology (triangles) compared with the
scaling relationship (equation~\ref{e:vexpand} with $\beta = 0.39$)
for an EdS background (crosses).  The two evolutions deviate by less
than 2\% at $\eta = 0.046$, the current age of the universe.}\label{evol}
\end{center}
\end{figure}

The addition of a cosmological constant ($\Lambda$) to the cosmology
is expected to slow down the conformal evolution of the voids at late
times.  We test this hypothesis by simulating the comoving evolution
of a single void of radius $25$ Mpc$/h$ starting at $z=1000$ in a
$(100$ Mpc$/h)^3$ box containing $64^3$ particles.  We find that the
void size evolution in a $\Lambda$CDM background deviates from that of
the EdS scaling solution at late times as expected.  However, the
final radius is underestimated by less than 2\% (see
figure~\ref{evol}).  The EdS void scaling relationship given by
equation~(\ref{e:vexpand}) with $\beta = 0.39$ is therefore a good
approximation for a void in a $\Lambda$CDM background.

\begin{figure}
\centering
\includegraphics[scale=0.3, angle=0]{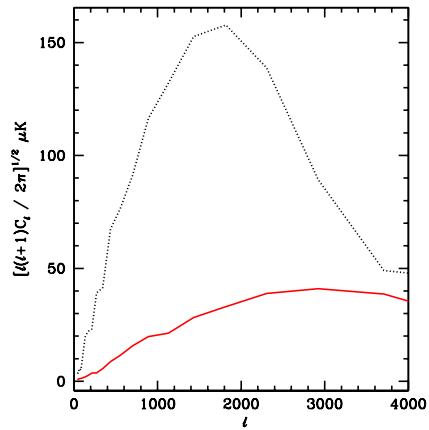}
\caption[equally sized voids]{\label{fig:lamveds} The predicted CMB
  power from a uniform distribution of equally sized voids in EdS
  (dotted) and $\Lambda$CDM (solid) background universes. The voids have a
  size of 40 Mpc$/h$ and a filling fraction of $40$ \%.  The void
  contribution is computed by ray tracing a void network.  The effect
  of $\Lambda$ is to suppress the effect of the voids and move the
  peak in the power spectrum to smaller angular scales since they will
  appear smaller on the last scattering surface.}
\end{figure}

Since a $\Lambda$CDM universe evolves for substantially longer, we
would expect the voids that we see today to appear smaller on the last
scattering surface for this cosmology than in our first approximation
EdS background. We would therefore predict that voids in a
$\Lambda$CDM universe will have a more suppressed effect on the CMB
than we have first approximated. This can be modelled by taking the
horizon size of our cone of voids to be that of a $\Lambda$CDM
cosmology. Figure~\ref{fig:lamveds} compares the predicted power from
a uniform distribution of equally sized voids in EdS and $\Lambda$CDM
background cosmologies.  The overall contribution to the CMB from the
voids is suppressed as expected and the peak is moved to smaller
angular scales since the voids now appear smaller on the last
scattering surface.

\subsection{Modelling acoustic oscillations}
So far we have studied voids which do not interact with their
surroundings. Specifically, they do not perturb the matter-radiation
fluid themselves. While this is a reasonable approximation for the RS
effect after recombination, it has been pointed out \citep{bac00} that
this is not true before last scattering. A void could therefore set up
oscillations in the matter-radiation fluid on scales up to the sound
horizon, which might be clearly visible in the power spectrum.

The standard way to include the full behaviour of matter and radiation
perturbations is to solve the Boltzmann equation numerically. But it
is not possible to model a void self-consistently inside the Boltzmann
solver. The dark matter density contrast inside the void becomes very
large (approaching $-1$ at late times) and is clearly
non-perturbative. A similar problem was encountered a few years ago
while investigating topological defects (see \cite{dur02} and
references therein). The solution in this case was to model the
defects as external sources. This can be done for the voids as
well. Any back reaction from the perturbation on the voids should only
be second order so can be neglected. Furthermore, the connection
between the external source (the voids) and the matter-radiation fluid
is via the metric perturbations $\Phi$ and $\Psi$ and, since they are
small during the times of interest, they can hence be handled within a
perturbative framework.

As opposed to defects, which are entirely external sources, the voids
represent a perturbation of the dark matter. By construction, the dark
matter is in a stable configuration and does not perturb itself. It is
therefore influenced by the presence of the void.  Also, since there
is no dark matter inside the void, there can be no flow outwards. We
have chosen to model these effects by making the approximation that
the dark matter is completely decoupled from the rest of the
universe. Therefore, the cold dark matter acts only as an external
seed where the perturbations are concerned (of course it is taken into
account for the background quantities).

We assume that perturbations driving away from the void configuration
are only due to the back-reaction from the induced perturbations in
the other fluids.  As these are second order effects, we can neglect
them. We could take them into account by coupling the dark matter to
the fluid perturbations alone. However, as expected, the effect on the
resulting power spectrum is minor, around 1\% for the scales of
interest which is comparable to the accuracy of the Boltzmann
codes. We prefer to use first order perturbation theory consistently
and turn the dark matter perturbations off completely.

We assume that the photons and baryons have flown relativistically
back into the void, while the cold dark matter remained frozen in the
wall. We therefore choose a density contrast
\begin{equation}
\delta = - \frac{\rho_c}{\rho_\gamma+\rho_\nu+\rho_b+\rho_c}
= -\frac{\Omega_c}{\Omega_{\rm rel}/a + \Omega_c + \Omega_b} \,.
\label{e:delta}
\end{equation}
In the phenomenological model investigated here, the voids emerge
slowly from the isotropic background of relativistic particles during
inflation.  Therefore, equation~\ref{e:delta} also describes the void
density contrast at high redshifts.  A more violent formation history
can lead to stronger acoustic waves and change our results.

Following the formalism of \citet{kun97}, \citet{dur97} and
\citet{dur99}, the source is found to be
\begin{eqnarray}
\Phi(\eta,{\mathbf k}) &=& -\frac{4\pi}{\sqrt{\eta_0^3}}
\left(\frac{a'}{a}\right)^2 \delta\, f(k,r_v(\eta))\\ f(k,r) &=&
\frac{\left( 3-(k r)^2\right) \sin(k r) - 3 k r \cos(k r)}{k^5} \,.
\label{e:four_source}
\end{eqnarray}
We insert this source into a modified version of CMBEasy
\citep{dor03}. We then write out a snapshot of the fluids at the time
of last scattering and Fourier transform them back.

\begin{figure}[t]
\begin{center}
\includegraphics[scale=0.3, angle=0]{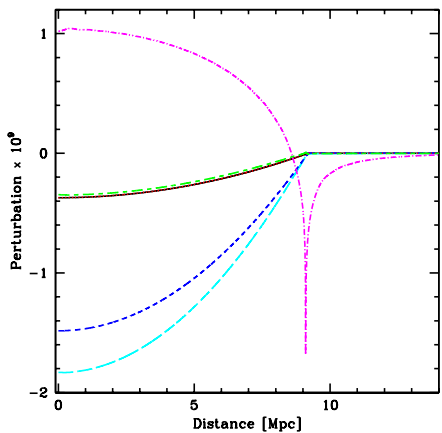}
\caption{The behaviour of the perturbations in real space at last
scattering (LS) from a single void. The void, which has a radius of 40
Mpc today, has a comoving size of 9 Mpc at LS. The black solid curve
is the input perturbation $\Phi_S(x,\eta_{LS})/2$, while the red
dotted curve is the total metric perturbation $\Phi$/2. The blue small
dashed curve shows $-2\Psi=2\Phi$. The cyan large dashed curve is the
temperature perturbation from the photons, $D_g/4$.  The baryon
velocity is shown as the magenta dot-dashed curve.  The green small
dashed-large dashed curve is $D_g/4-2 \Phi$, which is very similar to
$\Phi_S/2$.}\label{fig:tlss}
\end{center}
\end{figure}

\begin{figure}[t]
\begin{center}
\includegraphics[scale=0.3, angle=0]{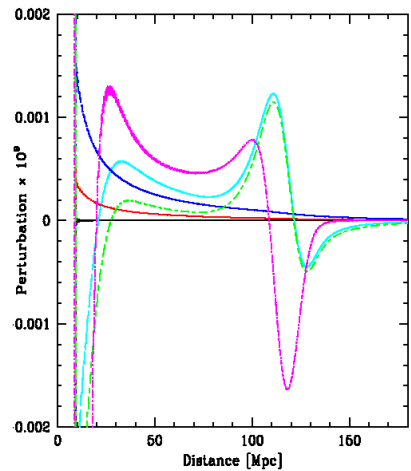}
\caption{A ``zoom'' onto the sound waves generated by the void. Notice
that the x axis now extends beyond the sound horizon (at about $111$
Mpc) and that the y axis has been magnified by roughly a factor 1000
compared to fig.~\ref{fig:tlss}. The lines are the same as in that
figure.}\label{fig:acoustic}
\end{center}
\end{figure}

Figure~\ref{fig:acoustic} shows a close up of the perturbations around
the sound horizon. These are due to oscillations set up during the
formation of the void and are therefore strongly dependent on the
formation history of the void. With our form of $\delta(\eta)$, given
by equation~(\ref{e:delta}), we see that the amplitude of the sound
waves is only about 1\% of the amplitude of the temperature
perturbations inside the void. For smaller voids, the relative
amplitude will be even less.  This seems quite unimportant, indeed,
the oscillations are not even visible in figure~\ref{fig:tlss}.
However, the void itself covers a surface of only about $9^2 \pi ({\rm
Mpc}/h)^2$ on the LSS, the sound horizon in our flat matter dominated
universe of the example case is at $111 {\rm Mpc}/h$ and the sound
waves extend out to about $100 {\rm Mpc}/h$, ten times further than
the size of the void. Therefore, the total power in the fluctuations
is about $100$ times larger than expected from their amplitude and is
comparable to the the power from the void predicted by our first
approximation method. Indeed, we see oscillations appear on large
angular scales in the angular power spectrum (see
figure~\ref{fig:compare}).
\subsection{Correlation function approach \label{sec:utc}}
                                                                                
In order to calculate the angular power spectrum of the CMB, we can again
use a technique developed for studying topological defects, namely the
use of unequal time correlators \cite{tur96}. The modified Boltzmann equation
described above is just a system of linear equations,
\begin{equation}
{\mathcal D} X = {\mathcal S} .
\end{equation}
Here ${\mathcal D}$ is a time dependent linear differential operator, $X$ is the
vector of perturbation variables and ${\mathcal S}$ the source term. For one single
void, the source is deterministic. But in general we are interested in
a void network with a random distribution of both the position and sizes
of the voids. In this case the source becomes a random variable.
                                                                                
For any given initial conditions we can formally solve the Boltzmann
equation by means of a Greens function ${\mathcal G}$,
\begin{equation}
X_j(t_0,{\mathbf k}) = \int_{t_{\rm in}}^{t_0} dt\,
{\mathcal G}_j(t_0,t,{\mathbf k}) S(t,{\mathbf k}) \,.
\end{equation}
The power spectrum is then a quadratic expectation value of the form
\begin{equation}
\begin{array}{l}
C_\ell \sim \left\langle X_j(t_0,{\mathbf k}) X_l^*(t_0,{\mathbf
k})\right\rangle = \\ \int_{t_{\rm in}}^{t_0} dt {\mathcal G}_j(t_0,t,{\mathbf
k}) \int_{t_{\rm in}}^{t_0} dt' {\mathcal G}_l(t_0,t',{\mathbf k}) \left\langle
{\mathcal S}(t,{\mathbf k}) {\mathcal S}^*(t',{\mathbf k})\right\rangle \, ,
\end{array}
\label{e:source}
\end{equation}
and we need to study the unequal time correlation function (UTC) of
the source in more detail,
\begin{equation}
\left\langle \Phi(t,{\mathbf k}) \Phi^*(t',{\mathbf k})\right\rangle .
\end{equation}
                                                                                
As the $\Phi$ of the different voids don't overlap in real space, we can
just add them up to create the total $\Phi$ of the void network,
\begin{eqnarray}
\Phi_{\rm tot} &=& \sum_j \Phi_j \\
&=& \sum_j \Phi^{(R_j)}({\mathbf x} - {\mathbf c}_j) .
\end{eqnarray}
The sum runs over all voids, $R_j$ is today's radius of void $j$ and
${\mathbf c}_j$ is the centre of void $j$. Going into Fourier space we find
\begin{eqnarray}
\Phi(t,{\mathbf k}) &=& \int d{\mathbf x} \Phi(t,{\mathbf x}) e^{i {\mathbf k} {\mathbf x}} \\
&=& \sum_j \int d{\mathbf x} \Phi^{(R_j)}(t,{\mathbf x}-{\mathbf c}_j) e^{i {\mathbf k} {\mathbf x}} \\
&=& \sum_j \Phi^{(R_j)}(t,{\mathbf k}) e^{i {\mathbf k} {\mathbf c}_j} \, .
\end{eqnarray}
The function $\Phi^{(R_j)}(t,{\mathbf k})$ is just the Fourier space source
for a single void at the origin with radius $R_j$ at time $t$ as
given in equation~(\ref{e:four_source}). The different spatial positions
of the voids are now encoded only in an additional phase factor.
The UTC required for computing the power spectrum is
\begin{equation}
\sum_{j,l} \Phi^{(R_j)}(t,{\mathbf k}) \Phi^{(R_l)}(t',{\mathbf k})
e^{i {\mathbf k} ({\mathbf c}_j-{\mathbf c}_l)} .
\label{e:utc}
\end{equation}
                                                                                
Let us first discuss the special case of equal-sized voids, $R_j
\equiv R\,\, \forall j$. In this case we can move the metric
perturbation $\Phi$ outside the sum and the UTC can be separated into
two parts,
\begin{equation}
\left\langle {\mathcal S}(t,{\mathbf k}) {\mathcal S}^*(t',{\mathbf k})\right\rangle .
= \Phi^{(R)}(t,{\mathbf k}) f({\mathbf k})\, \Phi^{(R)*}(t',{\mathbf k}) f({\mathbf k})
\end{equation}
where the $\Phi^{(R)}$ are again the single void metric
perturbations of equation~(\ref{e:four_source}) and
\begin{equation}
f({\mathbf k}) = \sqrt{\sum_{j,l} e^{i {\mathbf k} ({\mathbf c}_j-{\mathbf c}_l)}}
\end{equation}
The contribution to $f$ can be naturally separated into two
different parts, namely the contribution with $i=j$ which
amounts to $\sqrt{N_{\rm void}}$, the square root of the
number of voids in the universe, and the contribution with
$i\neq j$ which is a superposition of oscillations with
different frequencies, given by the distance between the
voids. These oscillations will in general tend to cancel out
so that we are left with $f({\mathbf k}) \approx \sqrt{N_{\rm void}}$.

The fact that the UTC separates into two parts is crucial,
since it permits us to integrate independently over $t$ and $t'$
in equation~(\ref{e:source}). Each integral amounts to the same as
solving the Boltzmann equation with the source
\begin{equation}
{\mathcal S}(t,{\mathbf k}) = \Phi^{(R)}(t,{\mathbf k}) f({\mathbf k})
\label{e:equal}
\end{equation}
and since they are the same, the angular power spectrum
is just the one computed with this source.
                                                                                
By setting $f({\mathbf k}) \approx \sqrt{N_{\rm void}}$ as discussed
above, and writing $C^{(R)}_\ell$ for the power spectrum which we
get with a single void as a source, we find
\begin{equation}
C_\ell \approx N_{\rm void} C^{(R)}_\ell
\label{e:eqsimple}
\end{equation}
and {\em not}
$C_\ell \approx N^2_{\rm void} C^{(R)}_\ell$ as might have been
guessed naively, the power spectrum being quadratic in $\delta T/T$.

In the general case of equation~\ref{e:utc} we cannot separate the UTC
into two parts. But now the oscillations in the $exp\{i {\mathbf k}
({\mathbf c}_j-{\mathbf c}_l)\}$ term are being helped to average out
by the now incoherent oscillations from the $\Phi^{(R_j)}$ with
different radii (as long as the void radii and the void positions are
not correlated). We therefore expect the main contribution to the
source to come from $i=j$, leaving us with
\begin{equation}
\sum_j \Phi^{(R_j)}(t,{\mathbf k}) \Phi^{(R_j)}(t',{\mathbf k})
\end{equation}
as the source. This is a sum of a source which separates into two
equal parts. Again writing $C^{(R)}_\ell$ for the single-void power
spectrum, we find
\begin{equation}
C_\ell = \sum_j C^{(R_j)}_\ell .
\end{equation}
This can be summed up using an actual realisation of a void distribution.
But since we are mainly interested in the average result for the theoretical
distribution, we can use this distribution directly. Writing $\rho(R) dR$
for the number of voids with radii between $R$ and $R+dR$ the total
power spectrum becomes
\begin{equation}
C_\ell = \int dR \rho(R) C^{(R)}_\ell .
\end{equation}

\subsection{Boltzmann imprint of primordial voids on the power spectrum}
\begin{figure}
\centering
\includegraphics[scale=0.3, angle=0]{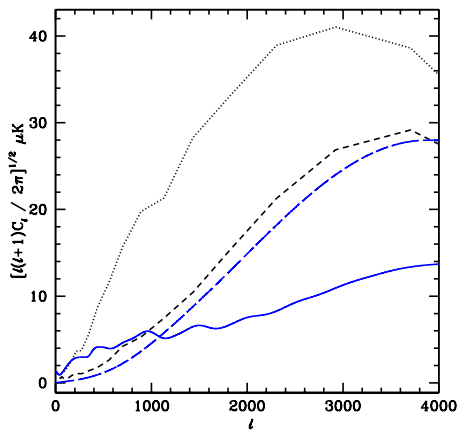}
\caption[equally sized voids]{\label{fig:compare} We compare the
  $C_\ell$ from ray tracing and the Boltzmann approach. The dotted
  curve shows the total $\delta T$ for a $\Lambda$CDM model with voids
  of a size of $40$ Mpc$/h$ and a filling fraction of $40$ \%, computed
  by ray tracing a void network. The solid curve shows the same but
  computed with a Boltzmann code. The small and large dashed lines show
  the RS effect due to $\Phi_v$ alone for both
  methods and with the same approximations, these two curves should coincide.}
\end{figure}

Figure~\ref{fig:compare} compares the void $C_\ell$ from ray tracing
and the Boltzmann approach for a uniform distribution of $40$ Mpc$/h$
sized voids.  The overall effect from the first approximation
ray tracing method appears to be suppressed using the improved
Boltzmann solution. Evidence of acoustic oscillations generated in the
photon-baryon fluid can also be seen on large angular scales. The RS
effect due to $\Phi_v$ alone, using the same approximations in both
calculations, is in agreement for both methods as expected.

The suppression that is evident using the Boltzmann solution over the
ray tracing approach is due to a number of factors.  In the ray tracing
method we assume that the density contrast $\delta=-1$ and we take
$(a'/a)^2$ to be $4/\eta^2$.  Close to radiation domination (ie. at
early times, just after last scattering, when we get the biggest
effect) both these approximations act to increase the ray tracing
result with respect to the more accurate Boltzmann prediction,
together by more than a factor of 2. The ray tracing method also
neglects the Doppler contribution from the baryon velocity, which
suppresses the result further.  Finally, the finite thickness of the
LSS and Silk damping were not fully taken into account in our first
approximation ray tracing method.

\begin{figure}[t]
\centering
\includegraphics[scale=0.3, angle=0]{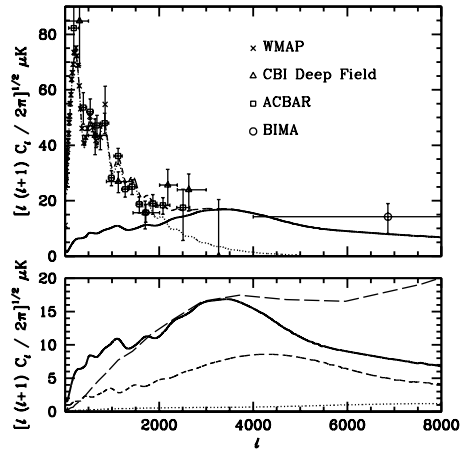}
\caption[equally sized voids]{\label{fig:final}Top: The CMB
anisotropies produced by a $\Lambda$CDM void model (solid line:
$\alpha = 3$, $r_{\rm max}=55$ Mpc$/h$ and $F_v=0.4$) compared
to the primary $\Lambda$CDM anisotropies (dotted). Also plotted are
the sum of primary and void contributions (dashed). We show the
``standard'' cosmological concordance model. Again a combined analysis
of primary and void--induced fluctuations would select a different
cosmology for the primary contribution.  Bottom: The large dashed
curve shows the fiducial EdS model of figure~\ref{figcl} ($\alpha =
3$, $r_{\rm max}=25$ Mpc$/h$ and $F_v=0.4$) using our ray
tracing method. The same model is shown for our Boltzmann solution
with EdS (small dashed) and $\Lambda$CDM (dotted) background
universes.  The solid bold line is the same $\Lambda$CDM void model
for $r_{\rm max}=55$ Mpc$/h$ maximally sized voids as in the
panel above.  }
\end{figure}

The parameter $r_{\rm max}$ is currently fixed using a number of void
analyses of surveys, including the 2-degree Field Galaxy Redshift
survey~\citep{hoy04}.  As discussed in subsection~\ref{sec:vstep},
these surveys only sample a fraction of the volume of the
universe. Therefore, the existence of voids of larger $r_{\rm max}$
than currently observed can not be ruled out. Models with $\alpha=3$
and high $r_{\rm max}$ voids in an EdS background cosmology tend to
predict too much power on scales $\ell \approx 1000$.  Our results
show that this is less of a problem for a $\Lambda$CDM
universe. Furthermore, since the Boltzmann solution predicts the CMB
power from voids to be lower than implied by the ray tracing
estimation, relatively large voids on the last scattering surface may
be consistent with the current CMB data, even for void distributions
with low values of $\alpha$ (see figure~\ref{fig:final}).

\section{Conclusions}
The cosmic microwave background is an excellent tool for probing the
distribution of matter from last scattering until today. In the case
of voids, the strongest signal stems from objects at very high
redshifts, especially from those already present at decoupling.  We
discuss in this Paper our improved method to investigate the imprint
of a distribution of primordial, spherical and compensated voids,
which could for example be generated by a phase transition during
inflation.

We show that the signature of a power-law distribution of such voids,
that is compatible with redshift sureys, contributes additional power
on small angular scales. The amount of power generated depends on the
parameters that describe the void size distribution as well as the
background cosmology. The overall signal predicted by our first
approximation ray tracing method is somewhat suppressed using our
improved Boltzmann solution. Thus, primordial void formation models
that produce relatively large voids on the last-scattering surface may
not be ruled out by the current CMB observations.

Experiments such as the CBI are able to directly probe small angular
scales and constrain void parameters. We will present a Markov Chain
Monte Carlo constraints analysis of a wide range of void models in a
future paper.  We will further constrain models that are compatible
with CMB observations using cluster evolution \citep{mat04} and also
investigate the non-Gaussian signal of void models that are compatible
with the observations.

Other sources are also expected to contribute at high $\ell$.
Probably the strongest of these is the thermal Sunyaev-Zel'dovich (SZ)
effect. Since the thermal SZ effect is strongly frequency--dependent,
experiments which work at about 30 GHz (like CBI) will see a stronger
signal than those working at higher frequencies \citep{agh02}. Hence a
multi-frequency approach should be able to easily disentangle the
contribution of voids from the SZ effect. Unfortunately, it seems to
be difficult at present to predict the precise level of the SZ
contribution, since different groups are reporting different results
(see eg. \citet{spr00} for a compilation).  Future multi-frequency,
high resolution and high signal-to-noise maps should be able to
significantly constrain the contribution of primordial voids to the
high $\ell$ CMB power spectrum.  Additionally, deep galaxy redshift
surveys and measurements of the distribution of matter in the
Ly-$\alpha$ forest will be able to directly explore the presence of
voids in the baryonic matter distribution at low redshifts.


\section*{Acknowledgments} 
We thank Michael Doran for a prerelease version of his CMBEasy code and
his support in modifying it. It is a pleasure to thank Ruth Durrer and
Andrew Liddle for helpful discussions and comments.  LMO acknowledges
support from ARC and PPARC.  MK acknowledges support by PPARC and the
Swiss Science Foundation.


\end{document}